\documentclass[fleqn,10pt]{wlscirep}
\title{Digital-analog quantum simulation of generalized Dicke models with superconducting circuits}

\author[1,*]{Lucas Lamata}
\affil[1]{Department of Physical Chemistry, University of the Basque Country UPV/EHU, Apartado  644, 48080 Bilbao, Spain}

\affil[*]{lucas.lamata@gmail.com}

\keywords{Quantum simulation, Superconducting circuits}

\begin{abstract}
We propose a digital-analog quantum simulation of generalized Dicke models with superconducting circuits, including Fermi-Bose condensates, biased and pulsed Dicke models, for all regimes of light-matter coupling. We encode these classes of problems in a set of superconducting qubits coupled with a bosonic mode implemented by a transmission line resonator. Via digital-analog techniques, an efficient quantum simulation can be performed in state-of-the-art circuit quantum electrodynamics platforms, by suitable decomposition into analog qubit-bosonic blocks and collective single-qubit pulses through digital steps. Moreover, just a single global analog block would be needed during the whole protocol in most of the cases, superimposed with fast periodic pulses to rotate and detune the qubits. Therefore, a large number of digital steps may be attained with this approach, providing a reduced digital error. Additionally, the number of gates per digital step does not grow with the number of qubits, rendering the simulation efficient. This strategy paves the way for the scalable digital-analog quantum simulation of many-body dynamics involving bosonic modes and spin degrees of freedom with superconducting circuits. 
\end{abstract}
\begin{document}

\flushbottom
\maketitle
\thispagestyle{empty}

\section*{Introduction}
The analysis of light-matter interactions has raised much interest in the past few years. Beginning with the seminal works by Rabi~\cite{Rabi36}, Dicke~\cite{Dicke54}, Jaynes and Cummings~\cite{JaynesCummings}, or Tavis and Cummings~\cite{TavisCummings}, current trends~\cite{Braak1,Braak2} aim at analyzing exotic regimes of light-matter coupling previously inaccesible to experimentation, including the ultrastrong coupling~\cite{Gunter09,Niemczyk10,FornDiaz10} and the deep-strong coupling~\cite{Casanova10,PolDSC,SembaDSC} regimes. While these scenarios are not achievable in standard atomic systems, the emergence of novel quantum technologies including superconducting circuits~\cite{Blais07} and semiconductor polaritons~\cite{Laussy} has recently enabled its exploration. Moreover, the field of quantum simulations~\cite{Feynman82,Lloyd96,Lanyon11,Salathe15,Barends15,Barends15a,Jane03,Nori,Nori2,Nori3,Buluta09,Georgescu14,Romero16} allows one to engineer light-matter interactions with a higher degree of controllability, and investigate these exotic coupling regions with enhanced accuracy and tunability. As some examples, an analog quantum simulation of the quantum Rabi model in all parameter regimes has been proposed in trapped ions~\cite{Pedernales15} and cold atoms~\cite{Felicetti16}. Additionally, a digital-analog version of the same model has been put forward in superconducting circuits~\cite{Mezzacapo14} with a preliminary analysis of Dicke physics, having been realized recently in the lab in the quantum Rabi case~\cite{Langford}. The digital-analog approach to quantum simulations (DAQS) promises to advance on the way to scalability, via the combination of large analog blocks with flexible digital steps. Further proposals for DAQS involve fermions~\cite{Casanova12,Mezzacapo12} and spins~\cite{Arrazola16} with trapped ions, fermion scattering with superconducting circuits~\cite{GarciaAlvarez15} and ions~\cite{Casanova11}, quantum chemistry with circuit quantum electrodynamics (QED)~\cite{GarciaAlvarez16}, and lattice gauge theories with cold atoms~\cite{Zohar16}.

The Dicke model~\cite{Dicke54} is one of the prototypical many-body models of light-matter coupling. It describes the interaction between a set of $N$ two-level atoms and a quantized bosonic mode of the electromagnetic field. In the classical spin limit, where $N$ goes to infinity, or the classical oscillator limit, where the ratio of the atomic transition frequency to the bosonic field frequency approaches infinity, the Dicke model is known to present a superradiant phase transition, for a certain value of the light-matter coupling~\cite{Lieb}. In many situations, a mean-field treatment of the dynamics suffices to capture the relevant physics. However, and given the recent interest in exploring the different regimes of light-matter coupling, a full-fledged quantum simulation of the Dicke model in all its parameter regimes, as well as with a variety of interactions involving inhomogeneities, biases, and time-dependent couplings, would be desirable.
In this sense, a first proof-of-principle demonstration of Dicke superradiance with two superconducting qubits in the weak coupling regime has been recently attained~\cite{Mlynek14}, and a pioneering experiment of the Dicke phase transition with a superfluid gas was performed~\cite{Baumann10}.

Here we propose the digital-analog quantum simulation of generalized Dicke models with superconducting circuits, which is a natural extension of Ref.~\cite{Mezzacapo14}. Besides estimating in detail the error in a previous scheme for the Dicke model with a DAQS~\cite{Mezzacapo14}, we propose a set of novel scenarios. These include the quantum simulation of Fermi-Bose condensates encoded onto broadband Dicke models~\cite{Altshuler}, biased Dicke models~\cite{HengFan,EmaryBrandesBiased}, as well as periodically-pulsed Dicke models~\cite{Dutta,BastidasPulsed}. After introducing the expressions for the respective quantum simulation dynamics, we obtain bounds and estimate the digital error analytically and numerically to validate the feasibility of the protocols for $N$ transmon qubits~\cite{Koch07,NoriFluxTransmon} coupled with a microwave bosonic mode, in a variety of parameter regions, from the weak coupling to the ultrastrong and deep-strong coupling regimes. This DAQS of the Dicke model and its generalizations are free from raised drawbacks presented by the square of the electromagnetic vector potential, $A^2$, in atomic systems or {\it ab initio} implementations of the Dicke model in superconducting circuits~\cite{Ciuti,Marquardt,Peropadre}. There is evidence that in these latter cases the superradiant phase transition will be absent, while with a DAQS the $A^2$ term effect may be in principle safely neglected. The digital-analog approach to quantum simulations may enable the study of mesoscopic many-body quantum systems in the near future, allowing us to progress towards the aimed-for quantum supremacy.

\section*{Results}

\subsection*{The Dicke model}
We consider a scenario with $N$ qubits coupled with a bosonic mode. This can be implemented with $N$ transmons~\cite{Koch07}  coupled to the electromagnetic field of a transmission line resonator.
The Hamiltonian for the Dicke model is given by ($\hbar=1$),
\begin{equation}
H_D=\sum_{i=1}^N\frac{\omega^D_0}{2} \sigma_z^i+\omega^D a^\dag a+\frac{\lambda^D}{\sqrt{N}}\sum_{i=1}^N \sigma_x^i (a+ a^\dag),
\end{equation}
where $\omega^D_0$($\omega^D$) is the frequency of the qubits(bosonic mode), $\lambda^D$ is the normalized coupling, $\sigma_k^i$, $k=x,z$ are Pauli operators,  and $a$ is the bosonic-mode annihilation operator. 

\begin{figure}[t]
\centering
\includegraphics[width=0.5\linewidth]{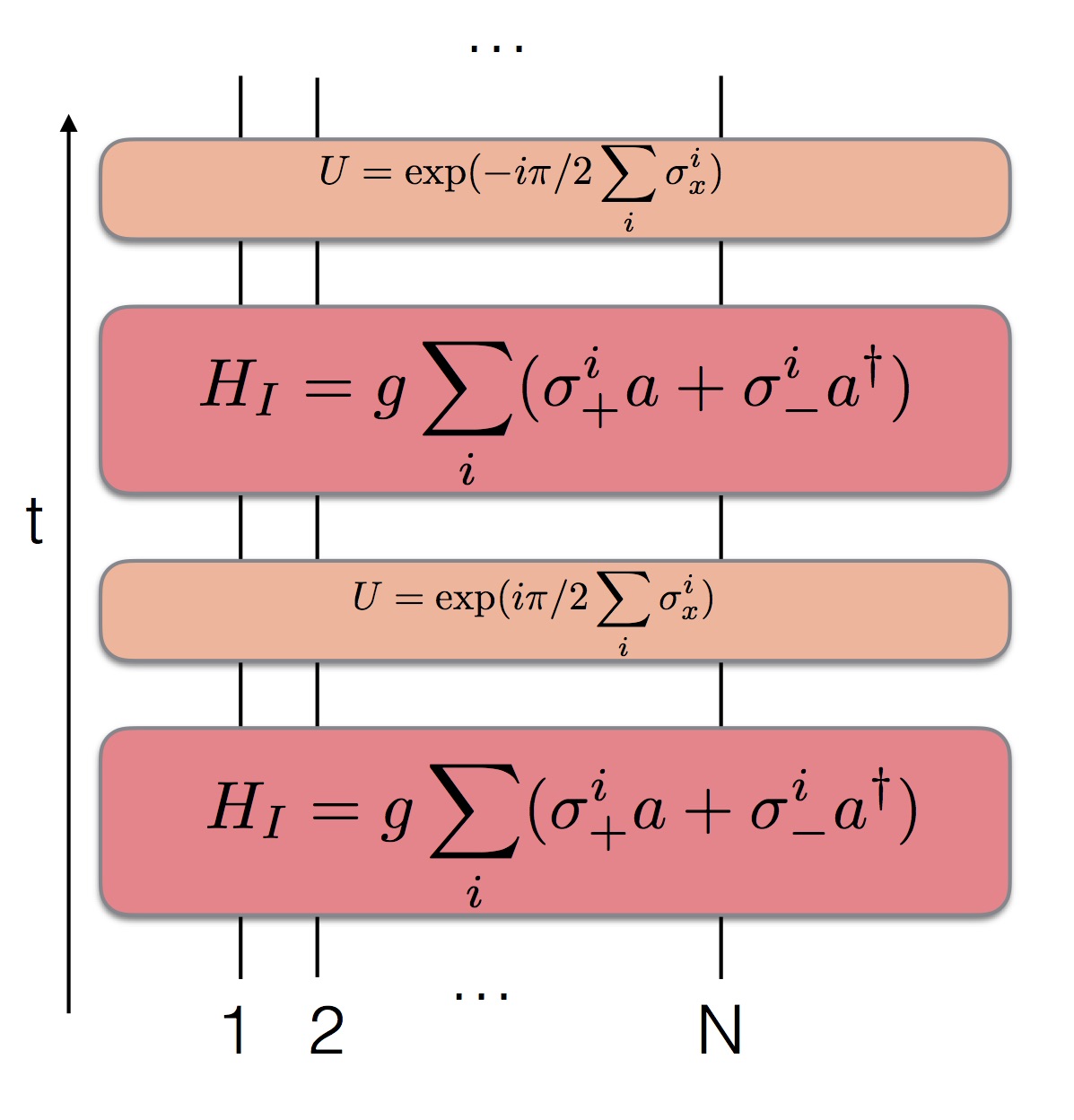}
\caption{\textbf{Digital-analog quantum simulation of the Dicke model.} We depict the circuit representation of a single Trotter step of the digital-analog approach to the quantum simulation of the Dicke model introduced in Ref.~\cite{Mezzacapo14}. The gate decomposition is based on iterated collective Tavis-Cummings dynamics, with two different alternative qubit detunings, interspersed with collective single-qubit rotations with respect to the $x$ axis. Interacting evolutions are depicted with red boxes, while single-qubit gates are plotted with orange boxes.}\label{fig:Fig1}
\end{figure}

The efficient DAQS of the Dicke model was schematized, without a detailed error analysis, by Mezzacapo et al.~\cite{Mezzacapo14}, via digitization of a combination of collective Tavis-Cummings interactions and collective single-qubit drivings, which generate anti-Tavis-Cummings interactions.  In order to implement the Dicke model and all its coupling regimes, one will first begin with the circuit QED Hamiltonian, in the strong-coupling regime, under rotating-wave approximation~\cite{Blais07},
\begin{equation}
H=\sum_{i=1}^N\frac{\omega_0}{2} \sigma_z^i+\omega a^\dag a+g\sum_{i=1}^N (\sigma_+^i a+ \sigma_-^i a^\dag),
\end{equation}
where $\omega_0(\omega)$ is the frequency of the transmons(transmission-line resonator), and $g$ is the transmon-resonator coupling strength.
We now move onto an interaction picture with respect to the free energy part  $\sum_i\frac{\delta}{2} \sigma_z^i+\delta a^\dag a$, which will produce the dynamics
\begin{equation}
H=\sum_{i=1}^N\frac{\omega_0-\delta}{2} \sigma_z^i+(\omega-\delta) a^\dag a+g\sum_{i=1}^N (\sigma_+^i a+ \sigma_-^i a^\dag).
\end{equation}
One will now perform a digitization of the dynamics, combining the previous collective Tavis-Cummings interaction with collective single-qubit rotations, changing the previous Hamiltonian onto an anti-Tavis-Cummings one,
\begin{equation}
\exp(i\pi/2\sum_i  \sigma_x^i) H \exp(-i\pi/2\sum_i  \sigma_x^i)=-\sum_{i=1}^N\frac{\tilde{\omega}_0-\delta}{2} \sigma_z^i+(\omega-\delta) a^\dag a+g\sum_{i=1}^N (\sigma_-^i a+ \sigma_+^i a^\dag),
\end{equation}
where the qubit detunings $\tilde{\omega}_0-\delta$ are now different to the previous case in order not to cancel the free energy part. The new frequency $\tilde{\omega}_0$ may take arbitrary values, while the simulating Dicke qubit frequency is related to the difference of $\omega_0$ and $\tilde{\omega}_0$ via $\omega_0^D=\omega_0-\tilde{\omega}_0$.

The way to combine both Hamiltonians, Tavis-Cummings and anti-Tavis-Cummings ones, together with the free energy terms, to obtain the Dicke model, is to employ the Lie-Trotter-Suzuki formula~\cite{Suzuki,Lloyd96}, which in its lowest order and for a Hamiltonian composed of two parts, $H=H_1+H_2$, corresponds to
\begin{equation}
e^{-iHt}\simeq (e^{-iH_1t/n}e^{-iH_2t/n})^n+O(t^2/n),
\end{equation}
where the error $O(t^2/n)$ depends on the commutator $[H_1,H_2]$ and decreases with the number of Trotter steps, $n$. In this approach, as already pointed out, $\omega_0^D=\omega_0-\tilde{\omega}_0$, the simulating Dicke bosonic mode frequency corresponds to $\omega^D=2(\omega-\delta)$, and the simulating Dicke coupling is $\lambda^D=\sqrt{N}g$. Below we include a detailed analysis of the contribution of the leading term to the digital error. We also include an analysis for the first time of the feasibility of the DAQS of the Dicke model with superconducting circuits, which was not carried out when the protocol was first introduced in Ref.~\cite{Mezzacapo14}. We obtain estimations of digital errors, Trotter steps, and gates, and compare them with state-of-the-art experiments of superconducting circuits with optimal coherence times.

Variants of the Dicke model can be considered, and will be analyzed below for the sake of versatile DAQS of generalized Dicke models. For example, one may have a finite qubit bandwidth, with different $\omega_0^i$ frequency for each qubit. This situation will give rise to the study of Fermi-Bose condensates~\cite{Altshuler}. One may also have a further  $\sum_{i=1}^N \Delta \sigma_x^i$ bias term, which will produce biased Dicke models~\cite{HengFan}. And finally, one may have a time-dependent $\lambda(t)$ coupling, related to periodically-pulsed Dicke models~\cite{Dutta}. All these situations are feasible to implement via digital-analog techniques in a setup composed of $N$ superconducting qubits coupled to a microwave resonator. Moreover, the flexibility of these techniques will allow one to explore all parameter regimes of the Dicke model and its generalizations, as well as getting rid of the detrimental effect of $A^2$ terms that may appear in direct implementations of the Dicke model in ultrastrong coupling circuit QED, which are not quantum simulations~\cite{Ciuti,Marquardt,Peropadre}. In the context of DAQS, the $A^2\propto (a+a^\dag)^2$ term will contain a renormalization of the mode frequency, which can be absorbed in $\omega$, and off-resonant terms, $a^2$, $(a^\dag)^2$, which will rotate fast in the strong-coupling regime here considered, and can be safely neglected. We point out that all the regimes of the Dicke model are here achieved via suitable transformation into an interaction picture, which cancels most of the $\omega_0^i$ and $\omega$ frequencies, and permits to tune their remaining values through appropriate qubit detunings. Therefore, in principle the $A^2$ term should not be an issue with a DAQS employing superconducting circuits.

We plot in Fig.~\ref{fig:Fig1} the circuit representation of a single Trotter step of the digital-analog approach to the quantum simulation of the Dicke model introduced in Ref.~\cite{Mezzacapo14}. The gate decomposition is based on iterated collective Tavis-Cummings dynamics, with two different alternative qubit detunings, interspersed with collective single-qubit rotations with respect to the $x$ axis. We point out that the number of gates per Trotter step does not grow with $N$, although each collective gate acts on all the $N$ qubits at the same time.

\begin{figure}[t]
\centering
\includegraphics[width=0.8\linewidth]{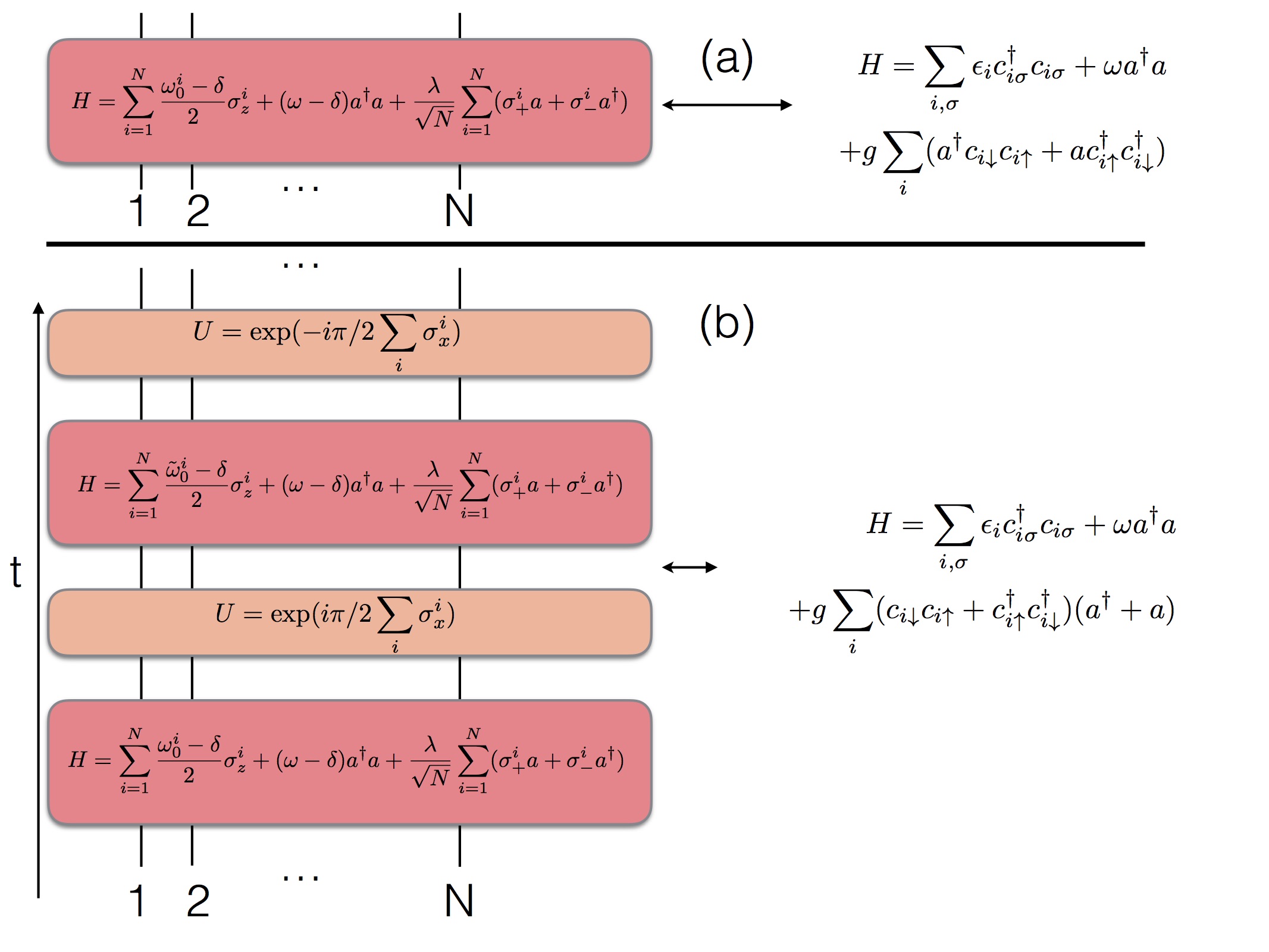}
\caption{\textbf{Digital-analog quantum simulation of a Fermi-Bose condensate via Dicke models.} (a) We depict the circuit representation of the fully analog approach to the quantum simulation of a Fermi-Bose condensate. The single Hamiltonian gives the full dynamics with no Trotterization. It is based on a Tavis-Cummings dynamics with inhomogeneous qubit detunings in a rotating frame. (b) A variant of the previous Fermi-Bose condensate with a modification of the pairing term can be obtained via a digital-analog implementation of a generalized Dicke model with inhomogeneous qubit detunings. The gate decomposition denotes a single Trotter step and is based on iterated collective Tavis-Cummings dynamics, with two different alternative qubit detunings, interspersed with collective single-qubit rotations with respect to the $x$ axis. Interacting evolutions are depicted with red boxes, while single-qubit gates are plotted with orange boxes.}\label{fig:Fig2}
\end{figure}

\subsection*{Fermi-Bose condensate}
The Hamiltonian for a Fermi-Bose condensate is given by~\cite{Altshuler}
\begin{equation}
H=\sum_{i,\sigma} \epsilon_i c^\dag_{i\sigma} c_{i\sigma} + \omega a^\dag a + g \sum_i (a^\dag c_{i\downarrow} c_{i \uparrow} + a c^\dag_{i\uparrow} c^\dag_{i\downarrow}),\label{eq1:FermiBose}
\end{equation}
where $\epsilon_i$ are the free energies of the fermionic modes, $\omega$ is the free energy of the bosonic mode, while $c_{i\sigma}(a)$ are annihilation fermionic(bosonic) operators, being $\sigma=\uparrow,\downarrow$ the fermionic spin. The dynamics given by this Hamiltonian can be retrieved with a classical computation in the mean-field approximation, when the number of bosonic excitations is large. However, here we focus in the exact behaviour, including all coupling regimes and valid for a wide range of bosonic excitations.

The Fermi-Bose condensate model of Eq.~(\ref{eq1:FermiBose}) can be mapped onto a generalized Dicke model~\cite{Altshuler}, where the qubit frequencies $\omega_0^i$ are inhomogeneous,
\begin{equation}
H=\sum_{i=1}^N\frac{\omega_0^i}{2} \sigma_z^i+\omega a^\dag a+\frac{\lambda}{\sqrt{N}}\sum_{i=1}^N (\sigma_+^i a+ \sigma_-^i a^\dag),\label{eq2:FermiBose}
\end{equation}
being $\sigma_z^i\equiv \sum_\sigma c^\dag_{i\sigma} c_{i\sigma}-I$, $\sigma_-^i\equiv c_{i\downarrow} c_{i \uparrow} $, and $\sigma_+^i\equiv c^\dag_{i\uparrow} c^\dag_{i\downarrow}$ Pauli operators. The mapping is exact in the unoccupied/double occupied subspace of levels $\epsilon_i$. The single-excitation subspace in each mode is decoupled from this dynamics via Eq.~(\ref{eq1:FermiBose}). The analogy is complete with $\epsilon_i= \omega_0^i/2$ and $g=\lambda/\sqrt{N}$. Eq.~(\ref{eq2:FermiBose}) is akin to a finite-bandwidth Tavis-Cummings model, which is the limit of the Dicke model in the coupling regime where $\lambda\ll \omega, \omega_0^i$. In this limit, the rotating-wave approximation holds, and the model in the zero-bandwidth case is analytically solvable.

The previous dynamics in Eq.~(\ref{eq2:FermiBose}) can be implemented with a purely analog quantum simulator using $N$ transmon qubits~\cite{Koch07}  with different detunings which are homogeneously coupled with the microwave electromagnetic field of a transmission line resonator. A variety of ranges for $\omega_0^i$, $\omega$, and $\lambda$ may be achieved via changing the detuning of the qubits and moving to an interaction picture with respect to part of the free energies, $\sum_i\frac{\delta}{2} \sigma_z^i+\delta a^\dag a$, which would leave as the new frame Hamiltonian the following,
\begin{equation}
H=\sum_{i=1}^N\frac{\omega_0^i-\delta}{2} \sigma_z^i+(\omega-\delta) a^\dag a+\frac{\lambda}{\sqrt{N}}\sum_{i=1}^N  (\sigma_+^i a+ \sigma_-^i a^\dag).\label{eq3:FermiBose}
\end{equation}
This can be straightforwardly implemented with current circuit QED technology~\cite{Blais07}.

On the other hand, more exotic Fermi-Bose Hamiltonians, of the form 
\begin{equation}
H=\sum_{i,\sigma} \epsilon_i c^\dag_{i\sigma} c_{i\sigma} + \omega a^\dag a + g \sum_i ( c_{i\downarrow} c_{i \uparrow} + c^\dag_{i\uparrow} c^\dag_{i\downarrow})(a^\dag+a),\label{eq3:FermiBose}
\end{equation}
correspond to the full-fledged generalized, broadband Dicke model without rotating-wave approximation, and can be implemented with a broadband modification of the introduced digital-analog protocol for the Dicke model, by modifying the qubit detunings to have different $\omega_0^i$ frequency for each qubit.

We plot in Fig.~\ref{fig:Fig2}a the circuit representation of the fully analog approach to the quantum simulation of the Fermi-Bose condensate described in Ref.~\cite{Altshuler}. The single Hamiltonian gives the full dynamics with no Trotterization. It is based on a Tavis-Cummings dynamics with inhomogeneous qubit detunings in a rotating frame. We depict in Fig.~\ref{fig:Fig2}b how a variant of the previous Fermi-Bose condensate with a modification of the pairing term can be obtained via a digital-analog implementation of a generalized Dicke model with inhomogeneous qubit detunings. The gate decomposition represents a single Trotter step involving iterated collective Tavis-Cummings dynamics, with two different alternative qubit detunings, interspersed with collective single-qubit rotations with respect to the $x$ axis.

\subsection*{Biased Dicke model}
A further variant of the Dicke model~\cite{HengFan} includes a bias term in the qubit free part, $\sum_{i=1}^N \Delta \sigma_x^i$, which sometimes appears in direct implementations of ultrastrong-coupling dynamics in superconducting circuits. Accordingly, the total Hamiltonian of this generalized Dicke model takes now the form, 
\begin{equation}
H=\sum_{i=1}^N\frac{\omega_0}{2} \sigma_z^i+\sum_{i=1}^N \Delta \sigma_x^i+\omega a^\dag a+\frac{\lambda}{\sqrt{N}}\sum_{i=1}^N \sigma_x^i (a+ a^\dag),
\end{equation}
where $\Delta$ is the bias frequency. This bias term is sometimes detrimental for the appearance of the superradiance phase transition, such that it is interesting to consider a DAQS that includes it to benchmark related physical behaviour. In order to implement this term and all the coupling regimes of this generalized Dicke model, one will first perform the Trotter step for the protocol of the Dicke model exposed above, and later add to it the necessary gates for obtaining the bias term, which may be implemented in different ways.  A possibility will come from additional digitization, via considering tunable-coupling qubits~\cite{Gambetta11,Houck11,MezzacapoMS} to cancel the interaction term, and $y$-rotation via collective microwave driving of the free energies of the qubits, to produce the $\sigma_x^i$ terms, as follows,
\begin{equation}
\exp(i\pi/4 \sum_i\sigma_y^i)\sum_i \sigma_z^i \exp(-i\pi/4 \sum_i \sigma_y^i)= \sum_i \sigma_x^i.
\end{equation}
This collective single-qubit rotation can be done straightforwardly via microwave pulses with auxiliary resonators addressing the transmon qubits.

\begin{figure}[t]
\centering
\includegraphics[width=0.5\linewidth]{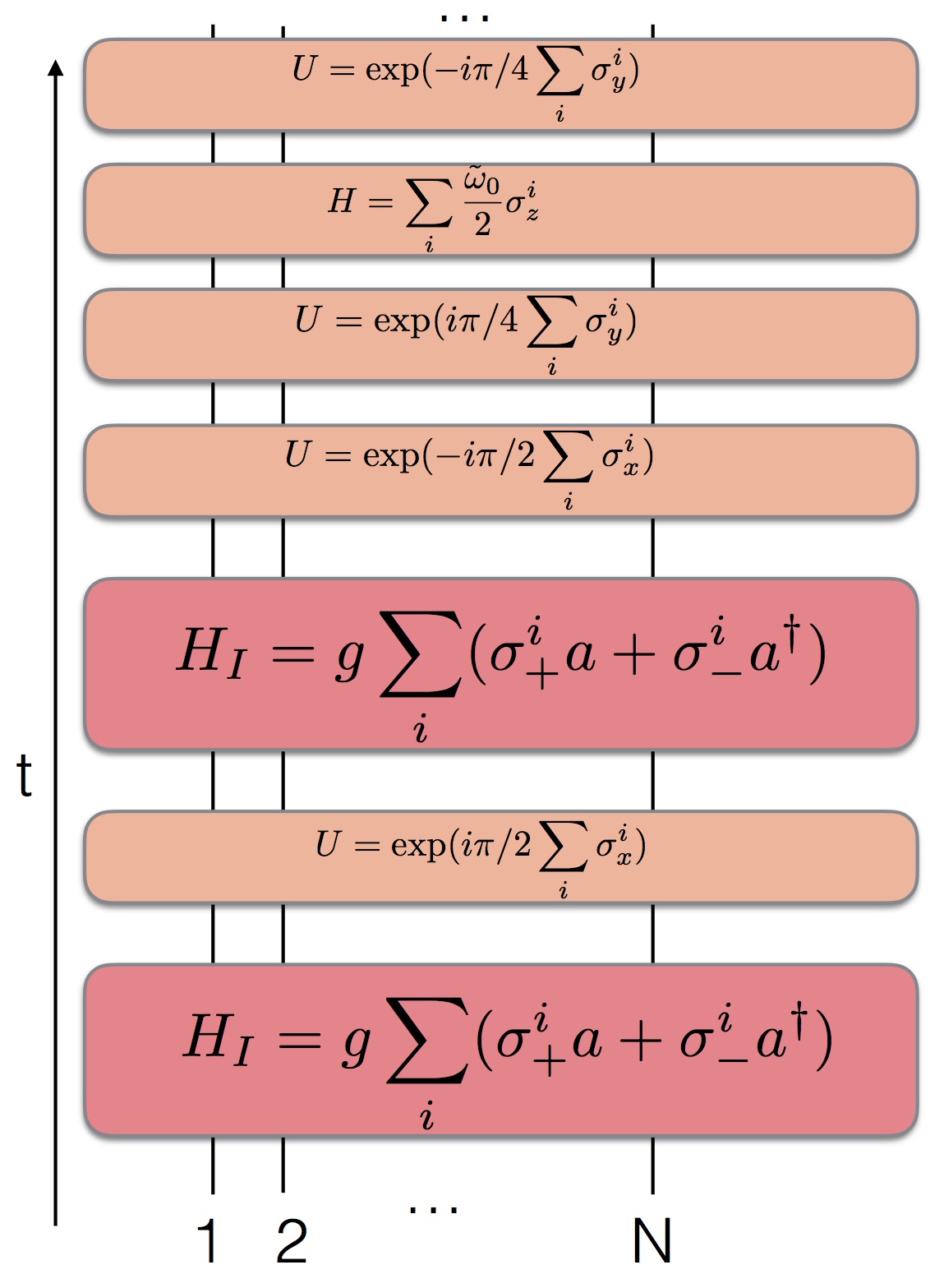}
\caption{\textbf{Digital-analog quantum simulation of a biased Dicke model.} We depict the circuit representation of a single Trotter step of the digital-analog approach to the quantum simulation of a biased Dicke model. The gate decomposition is based on iterated collective Tavis-Cummings dynamics, with two different alternative qubit detunings, interspersed with collective single-qubit rotations with respect to the $x$ axis, and a later evolution without qubit-resonator coupling, which rotates the qubits with collective single-qubit Y rotations onto the $x$ axis. Here the frequency $\tilde{\omega}_0$ of the simulating system is equal to $2\Delta$ in the simulated model.  Interacting evolutions are depicted with red boxes, while single-qubit gates and dynamics are plotted with orange boxes.}\label{fig:Fig3}
\end{figure}

We plot in Fig.~\ref{fig:Fig3} the circuit representation of a single Trotter step of the digital-analog approach to the quantum simulation of the biased Dicke model described in Ref.~\cite{HengFan}. This gate decomposition consists of iterated collective Tavis-Cummings dynamics, with two different alternative qubit detunings, interspersed with collective single-qubit rotations with respect to the $x$ axis, and a later evolution without qubit-resonator coupling, which rotates the qubits with collective single-qubit Y rotations onto the $x$ axis. 

\subsection*{Pulsed Dicke model}
Finally, the Hamiltonian for the periodically pulsed Dicke model~\cite{Dutta} is given by,
\begin{equation}
H=\sum_{i=1}^N\frac{\omega_0}{2} \sigma_z^i+\omega a^\dag a+\frac{\lambda(t)}{\sqrt{N}}\sum_{i=1}^N \sigma_x^i (a+ a^\dag),\label{eq1:pulsedDicke}
\end{equation}
where the dynamics now includes a kicked interaction term, with $\lambda(t)=\lambda_0+\lambda_1\sum_{k=1}^\infty \delta(t/T-2\pi k)$, for $\lambda_0$, $\lambda_1$ and $T$ constant parameters.

\begin{figure}[h]
\centering
\includegraphics[width=0.5\linewidth]{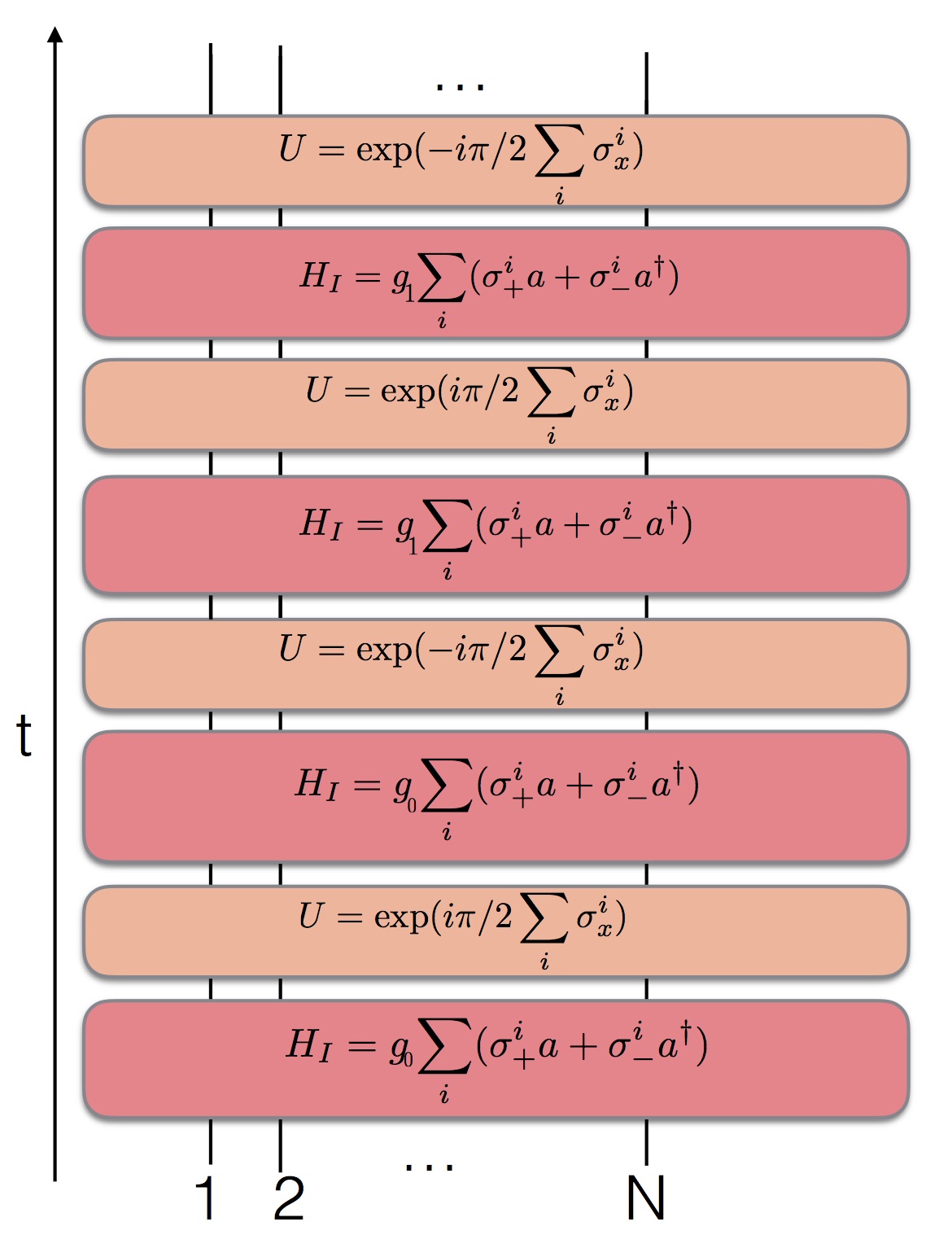}
\caption{\textbf{Digital-analog quantum simulation of a pulsed Dicke model.} We depict the circuit representation of a single Trotter step of the digital-analog approach to the quantum simulation of a periodically-pulsed Dicke model. The gate decomposition is based on two iterated collective Tavis-Cummings dynamics with different couplings, $g_0$, $g_1$, each of them with two different alternative qubit detunings, interspersed with collective single-qubit rotations with respect to the $x$ axis. Interacting evolutions are depicted with red boxes, while single-qubit gates are plotted with orange boxes.}\label{fig:Fig4}
\end{figure}

The dynamics of this periodically pulsed Dicke model can be analyzed with Floquet theory and includes a shift in the quantum critical point, as well as the appearance of sideband quantum phase transitions and novel phases~\cite{Dutta}.

The digital-analog approach to quantum simulation is very appropriate for this kind of scenario.
One may straightforwardly implement this generalized dynamics via tunable-coupling transmon qubits~\cite{Gambetta11,Houck11,MezzacapoMS}, considering two values of the couplings, $\lambda_0/\sqrt{N}$ and $(\lambda_0+\lambda_1 \alpha)/\sqrt{N}$, where we approximate the Dirac delta pulses by square pulses of height $\alpha$ and width $\tau$, which in the limit $\alpha\rightarrow\infty$, $\tau\rightarrow 0$, $\alpha \tau=1$, will recover the delta pulses. We point out that this limit does not increase the digital error given that the quantity that enters into the commutator of the different terms, contributing to the error, depends on the product $\alpha\tau=1$, which does not explode as $\alpha$ grows. Moreover, by fast switching of the coupling via external magnetic fluxes, as well as fast single-qubit rotations, a large amount of digital steps may be produced, reducing the digital error. Additionally, the same physics as for the kicked Delta pulses is recovered with square pulses under certain limits~\cite{Dutta}, enabling the DAQS.  

The protocol for this case will begin with the circuit QED Hamiltonian, in the same interaction picture as in previous cases, 
\begin{equation}
H=\sum_{i=1}^N\frac{\omega_0-\delta}{2} \sigma_z^i+(\omega-\delta) a^\dag a+g(t)\sum_{i=1}^N (\sigma_+^i a+ \sigma_-^i a^\dag),
\end{equation} 
where $g(t)$ will be a bimodal function taking values $g_0=\lambda_0/\sqrt{N}$, and $g_1=(\lambda_0+\lambda_1\alpha)/\sqrt{N}$, and where the value of $g(t)$ can be tuned via the structure of the tunable-coupling transmon qubit, with the application of external magnetic-flux drivings.

The protocol will proceed similarly as in the purely Dicke case, except that now, inside a Trotter step, instead of a single-coupling Tavis-Cummings and anti-Tavis-Cummings dynamics, we will have two different Dicke models subsequently implemented, each of them with each of the $g_{0,1}$ couplings, which will be performed each with a pair of Tavis-Cummings and anti-Tavis-Cummings interactions. In the limit of large number of Trotter steps, $n\gg 1$, the dynamics of Eq.~(\ref{eq1:pulsedDicke}) will be recovered.

We plot in Fig.~\ref{fig:Fig4} the circuit representation of a single Trotter step of the digital-analog approach to the quantum simulation of the periodically-pulsed Dicke model described in Ref.~\cite{Dutta}. This gate decomposition relies on two iterated collective Tavis-Cummings dynamics with different couplings, $g_0$, $g_1$, each of them with two different alternative qubit detunings, interspersed with collective single-qubit rotations with respect to the $x$ axis. 

\subsection*{Estimation of the digital error}
The lowest order contribution to the digital error is given by the expression~\cite{Lloyd96}
\begin{equation}
\epsilon(H_i)=\sum_{i< j}\frac{[H_i,H_j] t^2}{2n},\label{eqmain:digitalerror}
\end{equation}
while the next orders go as $O(t^3/n^2)$ and often may be neglected for large $n$.

In this section we estimate the contribution to the error produced by the leading term in digital-analog implementations of the Dicke model and its generalizations. Therefore, we compute the different commutators of the various parts involved, namely, in the case of the Dicke model,
\begin{eqnarray}
H_1=\sum_i \frac{\omega_{0,1}^i}{2} \sigma_z^i+\tilde{\omega} a^\dag a + g \sum_i  (\sigma_+^i a+ \sigma_-^i a^\dag),
\end{eqnarray}
and 
\begin{eqnarray}
H_2=\sum_i \frac{\omega_{0,2}^i}{2} \sigma_z^i+\tilde{\omega} a^\dag a + g \sum_i  (\sigma_-^i a+ \sigma_+^i a^\dag),
\end{eqnarray}
where $\omega^i_{0,k}$, $k=1,2$ are the effective qubit frequencies which appear in the interaction picture, and the effective mode frequency $\tilde{\omega}$ is also assumed to be computed in the interaction picture.

One has then that
\begin{equation}
[\sum_i(\sigma_+^i a+ \sigma_-^i a^\dag ),\sum_i( \sigma_-^i a + \sigma_+^i a^\dag)]=\sum_i \sigma_z^i [a^2-(a^\dag)^2]+\sum_{ij} (\sigma_+^i\sigma_+^j-\sigma_-^j \sigma_-^i),
\end{equation}
while
\begin{equation}
[\sum_i(\sigma_+^i a+ \sigma_-^i a^\dag ),\sum_i \sigma_z^i]=2 \sum_i(\sigma_-^i a^\dag-\sigma_+^i a),
\end{equation}
and
\begin{equation}
[\sum_i(\sigma_+^i a+ \sigma_-^i a^\dag ),a^\dag a]=\sum_i (\sigma_+^i a-\sigma_-^i a^\dag).
\end{equation}

Therefore,
\begin{eqnarray}
\epsilon(H_1,H_2) & = & [\sum_i g \omega_{0,1}^i t^2 (\sigma_+^i a^\dag-\sigma_-^i a) + \sum_i g \tilde{\omega} t^2 (\sigma_+^i a^\dag-\sigma_-^i a)+ \sum_i g \omega_{0,2}^i t^2 (\sigma_-^i a^\dag-\sigma_+^i a) + \sum_i g \tilde{\omega} t^2 (\sigma_+^i a-\sigma_-^i a^\dag)\nonumber \\&&+ (gt)^2 \{\sum_i \sigma_z^i [a^2-(a^\dag)^2]+\sum_{ij} (\sigma_+^i\sigma_+^j-\sigma_-^j \sigma_-^i)\}]/2n.\label{eq1:DigitalError}
\end{eqnarray}
For a sizable dynamics, we may assume that $\max\{g,\omega_{0,k}^i,\tilde{\omega}\} t \simeq 1$, and give a bound on the leading contribution to the error via the Cauchy-Schwarz inequality,
\begin{eqnarray}
||\epsilon(H_1,H_2)||\lesssim [4 N (||a||+||a^\dag||) + N ||a^2-(a^\dag)^2|| +N^2]/2n,\label{eq:looseboundDigitalError}
\end{eqnarray}
where we consider the supremum norm over the involved states in the dynamics, for which $||\sigma_k||=1$, with $k=x,y,z$.Here, in order to compute the norm of the unbounded operators $a$ and $a^\dag$, we consider them to be restricted to a finite-dimensional domain Hilbert space, such that this set will include all the appearing states in the considered evolution.

Accordingly, and assuming that the number of bosonic excitations remains bounded during the evolution, the scaling of the leading error contribution in the number of qubits $N$ of the implementation is at most quadratic, as one would expect. We point out that the tail of bosonic excitations can be long, i.e., populating many Fock states, but with a low probability, such that classically the computation will be hard, having to truncate at a high dimension, but for a DAQS the norm of the domain-restricted $a$ operators needs not be large for this to happen. For example, this will take place for a small average value and a large variance of $a^\dag a$. Moreover, as pointed out above, each digital step in the digital-analog implementation of the Dicke model is composed of two Tavis-Cummings interactions and two collective single-qubit rotations. Therefore, the number of gates does not grow with $N$, which is a nice feature of this implementation. Below we include numerical simulations benchmarking the DAQS of the Dicke model and the pulsed Dicke model with superconducting circuits, including the digital error as well as realistic decoherence sources. The previous bound in the error is intended to analyze its scaling with the number of qubits. The final digital error will depend on how many bosonic excitations will be produced, which will strongly depend on the particular cases involved. In many situations, the bosonic excitations grow until a certain value and then decrease, as happens in the quantum Rabi model with finite mode frequency, see Ref.~\cite{Casanova10}, or in the inhomogeneous Dicke model wherever closed subspaces are produced in the dynamics, see Ref.~\cite{Peng16}. The numerical simulations we performed show that the protocol is feasible for the analyzed cases, where the number of bosonic excitations does not grow much. Additionally, the recent implementation of a digital-analog quantum simulator of the quantum Rabi model~\cite{Langford} based on the proposal by Mezzacapo et al.~\cite{Mezzacapo14}, achieves a small digital error even for a significantly high number of about 30 bosonic excitations.

\begin{figure}[ht]
\centering
\includegraphics[width=0.6\linewidth]{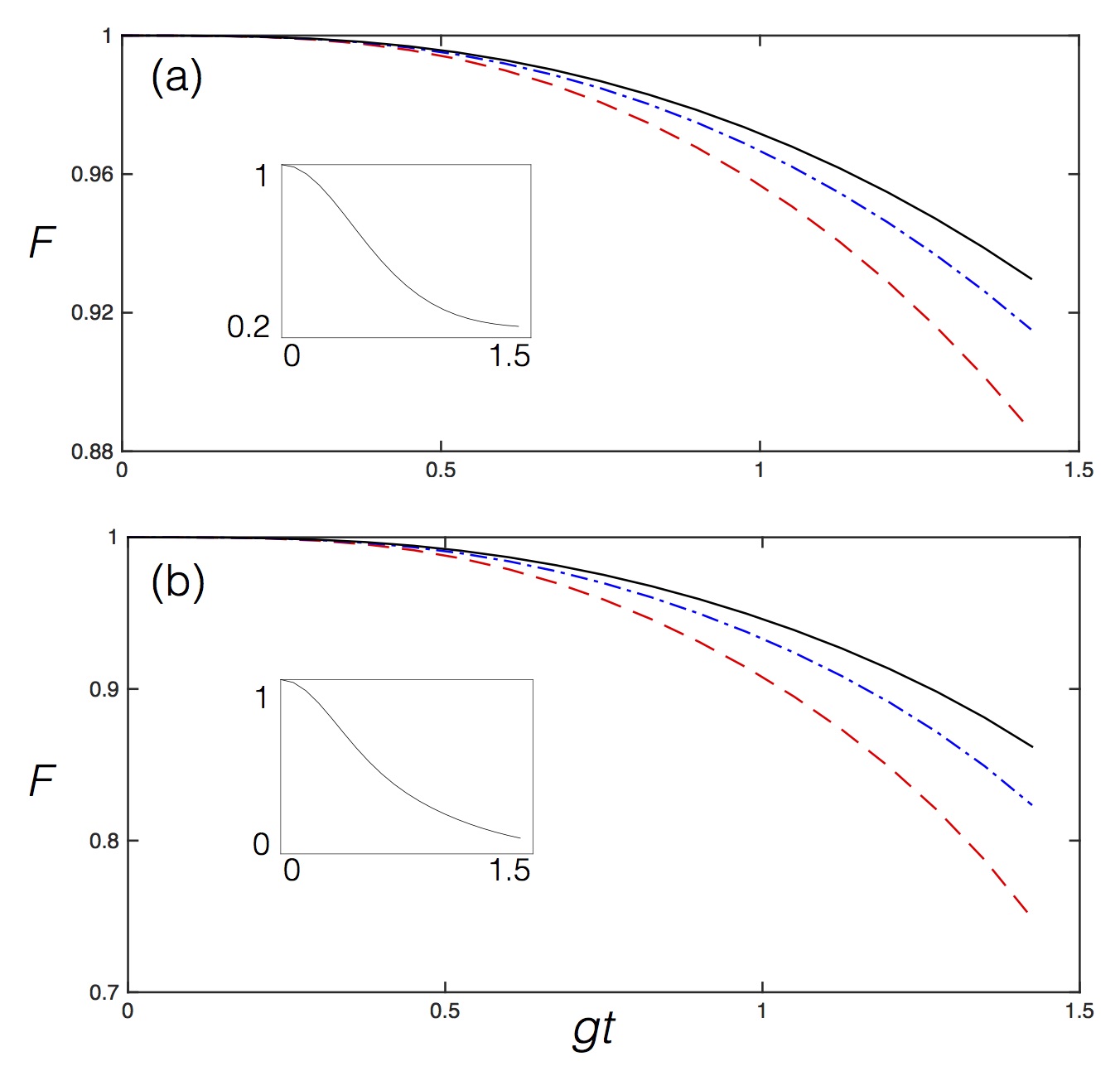}
\caption{\textbf{Dicke model. Fidelity $F$ for (a) $N=2$ and (b) $N=3$, $n=7,9,11$: Deep-strong coupling.} Numerical simulation with realistic decoherence sources of the implementation with circuit QED of the Dicke model for (a) $N=2$ and (b) $N=3$, with $n=7$ (dashed), $n=9$ (dashed-dotted) and $n=11$ (solid) Trotter steps, for a  coupling/resonator ratio $\lambda/\sqrt{N}\omega=1.5$, where $\omega$ denotes the resonator frequency of the simulated model, and a qubit to resonator frequency ratio $\omega_0/\omega=1/20$, where $\omega_0$ denotes the qubit frequency of the simulated model. Decoherence sources include that of the resonator $\kappa$, with $\kappa/\omega=10^{-2}$, spontaneous emission $\Gamma_s$, and dephasing $\Gamma_d$, with $\Gamma_s/\omega=\Gamma_d/\omega=0.5\times 10^{-2}$. We plot the fidelity $F={\rm Tr}(\rho_T \rho_I)$ of the overlap between the states $\rho_T$ of the Trotterized evolved dynamics including decoherence sources, and the ideal evolution $\rho_I$, versus $gt$, with $g=\lambda/\sqrt{N}$. We depict in the inset the survival probability of the initial state under ideal evolution, showing that the dynamics in the considered time interval is sizable, namely, the state is significantly evolved.}\label{fig:Fig6}
\end{figure}

A similar consideration of the digital error can be done with the Fermi-Bose condensate, given that the single Trotter step has a related structure to the pure Dicke model. The only difference appears in the inhomogeneous qubit frequencies in the former, but a similar scaling can be expected for a wide range of parameter values.

Regarding the biased Dicke model, there appears a third term in the Hamiltonian, $H_0=\sum_i \Delta \sigma_x^i$, which contributes to the leading term of the digital error via the commutators with $H_1$ and $H_2$, as shown in Eq.~(\ref{eqmain:digitalerror}). Accordingly, the following expression results for the equivalent bound to Eq.~(\ref{eq1:DigitalError}),
\begin{eqnarray}
\epsilon_{\rm Bias}(H_0,H_1,H_2)=\epsilon(H_1,H_2)-i\sum_i \Delta t^2\frac{(\omega_{0,1}^i+\omega_{0,2}^i)}{2n} \sigma_y^i,
\end{eqnarray}
contributing an additional term linear in $N$ to Eq.~(\ref{eq:looseboundDigitalError}),
\begin{equation}
||\epsilon_{\rm Bias}(H_0,H_1,H_2)||\lesssim ||\epsilon(H_1,H_2)|| + N/n.
\end{equation}
We point out that the new term is subdominant with respect to the quadratic term, and for large number of qubits, in many situations, will not contribute much to the digital error.

Finally, in the periodically pulsed Dicke model, the analysis amounts to duplicating the number of gates per Trotter step as compared with the Dicke model. This is approximately equivalent to duplicating the number of necessary Trotter steps for achieving the same fidelity, given that the gates are formally equivalent to this case except by the couplings of the interactions. Therefore, in this case $\epsilon_{\rm Pulse}(H_i)\simeq 2\epsilon(H_1,H_2)$.

Current experiments with superconducting circuits have reached more than 1000 gates in a digital quantum simulator, and decoherence times are about three orders of magnitude longer than gate times~\cite{Barends15a} such that one may expect that a proof-of-principle experiment of the Dicke model in all its parameter regimes, from the weak to the ultrastrong and deep-strong coupling regimes, may be carried out with state-of-the-art or near future technology. Further improvements in gate fidelities and coherence times may allow one to reach many tens of qubits with many bosonic excitations in a full-fledged Dicke or generalized Dicke quantum simulator, such that this way one may reach quantum supremacy without the need of quantum error correction.

\subsection*{Numerical simulations}
We have performed numerical simulations for the DAQS of the Dicke model and pulsed Dicke model with master equation and realistic decoherence sources for superconducting circuits, to show the feasibility of the protocol. We Trotterize the dynamics and compare the ideal, exact evolution with the digitized evolution in the presence of decoherence. For including decoherence we employed a master equation formalism with a unitary part based on the corresponding Dicke or pulsed Dicke Hamiltonian, and a Lindblad part composed of cavity damping, spontaneous emission, and dephasing, following the expression,
\begin{eqnarray}
\frac{d\rho}{dt}=-i[H,\rho]+\kappa(2a\rho a^\dag-a^\dag a\rho-\rho a^\dag a)/2+\Gamma_s \sum_i(2\sigma_-^i\rho\sigma_+^i-\sigma_+^i\sigma_-^i\rho-\rho\sigma_+^i\sigma_-^i)/2+\Gamma_d \sum_i(\sigma_z^i\rho\sigma_z^i-\rho).\label{LindbladTrotter}
\end{eqnarray}
To implement the digital-analog dynamics, we apply successively the evolution of Eq.~(\ref{LindbladTrotter}) for each Trotter step component, employing the corresponding Hamiltonian, e.g., Tavis-Cummings and local gates.

We plot in Fig.~\ref{fig:Fig6} the fidelity $F= {\rm Tr}(\rho_T \rho_I)$ for (a) $N=2$ and (b) $N=3$, of the overlap between the states $\rho_T$ of the Trotterized evolved dynamics including decoherence sources, and the ideal evolution $\rho_I$, versus $gt$, with $g=\lambda/\sqrt{N}$. Here $g/\omega=1.5$. This case corresponds to the deep-strong coupling regime. We depict in the inset the survival probability of the initial state, which is the ground state of qubits and bosonic mode of the free Hamiltonian, showing that this state significantly changes during this dynamics. We observe that this state is not preserved as corresponds to this coupling/mode frequency ratio, which includes the counterrotating terms of the Dicke Hamiltonian.

\begin{figure}[t]
\centering
\includegraphics[width=0.7\linewidth]{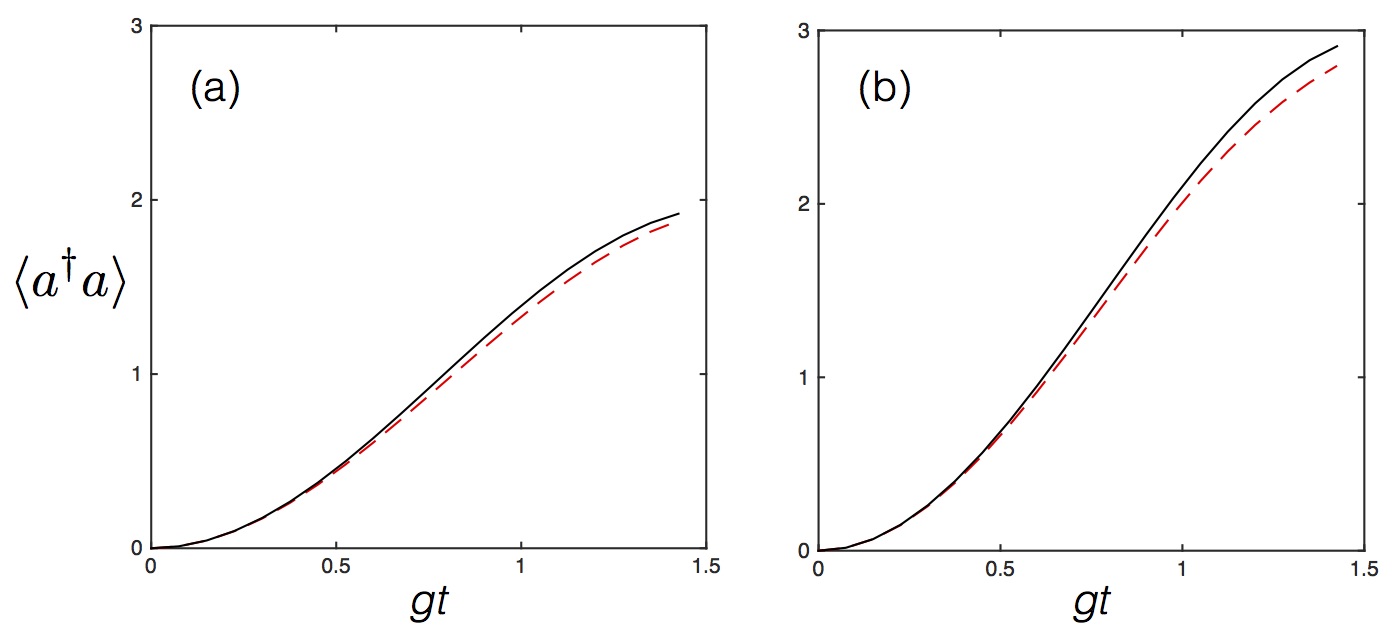}
\caption{\textbf{Dicke model. Photon number average for (a) $N=2$ and (b) $N=3$, $n=7$: Ultrastrong coupling.} Numerical simulation with realistic decoherence sources of the implementation with circuit QED of the Dicke model for (a) $N=2$ and (b) $N=3$ with $n=7$ Trotter steps, for a  coupling/resonator ratio $\lambda/\sqrt{N}\omega=0.5$, where $\omega$ denotes the resonator frequency of the simulated model, and a qubit to resonator frequency ratio $\omega_0/\omega=1/20$, where $\omega_0$ denotes the qubit frequency of the simulated model. Decoherence sources include that of the resonator $\kappa$, with $\kappa/\omega=10^{-2}$, spontaneous emission $\Gamma_s$, and dephasing $\Gamma_d$, with $\Gamma_s/\omega=\Gamma_d/\omega=0.5\times 10^{-2}$. We plot the photon number average $\langle a^\dag a\rangle={\rm Tr}(a^\dag a\rho_{T,I})$ with respect to the states $\rho_T$ (dashed) of the Trotterized evolved dynamics including decoherence sources, and the ideal evolution $\rho_I$ (solid), versus $gt$, with $g=\lambda/\sqrt{N}$.}\label{fig:Fig8}
\end{figure}

One can appreciate that the fidelity $F$ decreases from the $N=2$ to the $N=3$ cases, as one would expect due to larger digital error as well as increased imperfections due to the larger size. Nevertheless, for a still reduced number of Trotter steps of 11, as compared with the already experimentally achieved numbers of about 90, see Langford et al.~\cite{Langford}, the fidelity remains large with respect to the classical limit for a randomly chosen state, during a long time evolution in which the initial state changes significantly, namely, for a sizable dynamics.

\begin{figure}[ht]
\centering
\includegraphics[width=0.6\linewidth]{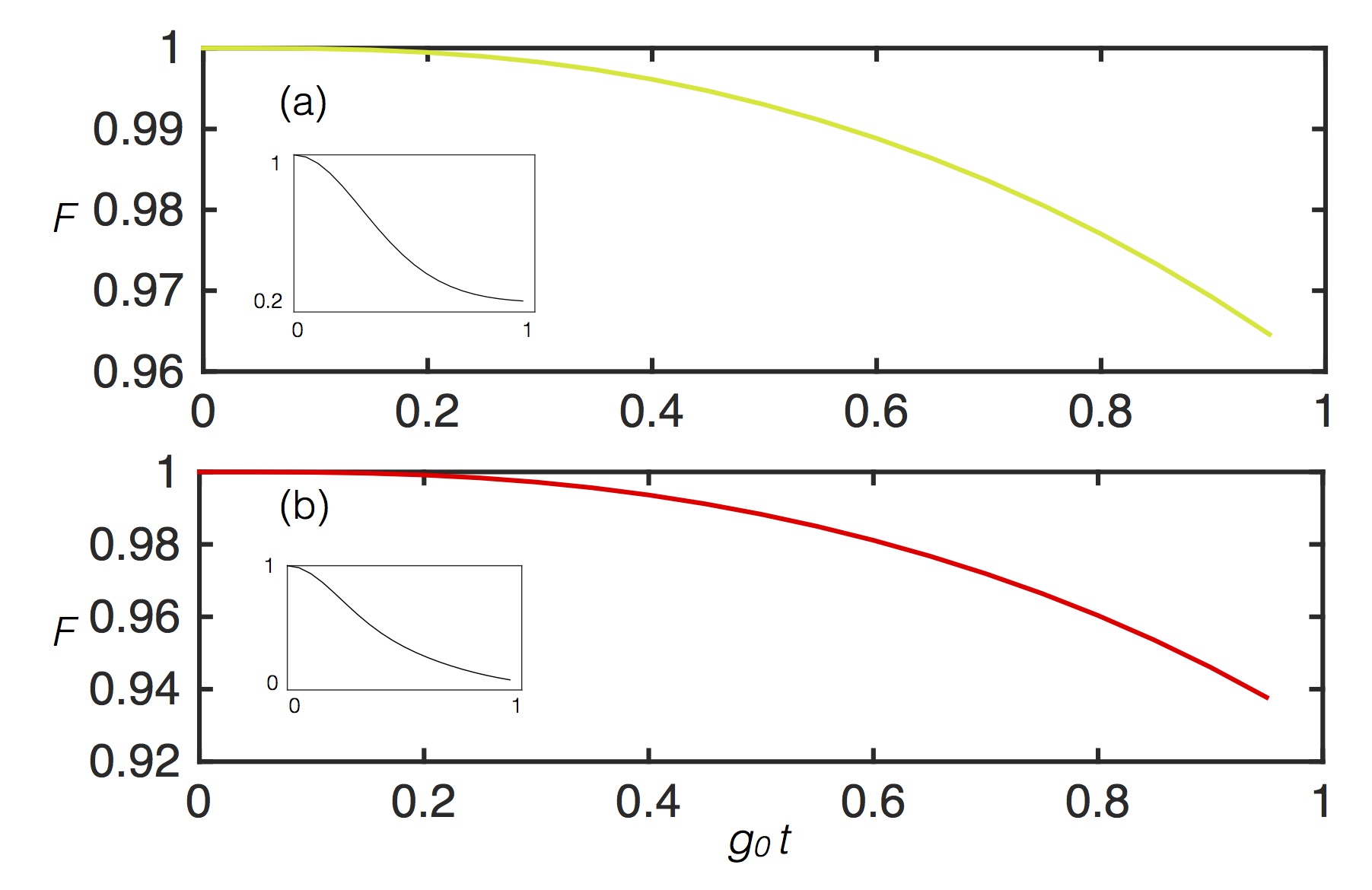}
\caption{\textbf{Pulsed Dicke model. Fidelity $F$ for (a) $N=2$ and (b) $N=3$, $n=13$: Deep-strong coupling.} Numerical simulation with realistic decoherence sources of the implementation with circuit QED of the pulsed Dicke model for (a) $N=2$ and (b) $N=3$, with $n=13$ Trotter steps, for a  coupling/resonator ratio $g_0/\omega=1.5$ and $g_1=2 g_0$, where $\omega$ denotes the resonator frequency of the simulated model, and a qubit to resonator frequency ratio $\omega_0/\omega=1/20$, where $\omega_0$ denotes the qubit frequency of the simulated model. In this specific simulation we made the periodicity of the Trotter steps coincide with the one of alternating couplings $g_0$ and $g_1$, as indicated in Fig.~\ref{fig:Fig4}.  The time for evolution with $g_0$ and $g_1$ couplings is taken to be equal for both. Decoherence sources include that of the resonator $\kappa$, with $\kappa/\omega=10^{-2}$, spontaneous emission $\Gamma_s$, and dephasing $\Gamma_d$, with $\Gamma_s/\omega=\Gamma_d/\omega=0.5\times 10^{-2}$. We plot the fidelity $F={\rm Tr}(\rho_T \rho_I)$ of the overlap between the states $\rho_T$ of the Trotterized evolved dynamics including decoherence sources, and the ideal evolution $\rho_I$, versus $g_0t$, with $g_0=\lambda_0/\sqrt{N}$ and $g_1=(\lambda_0+\lambda_1 \alpha)/\sqrt{N}$. We depict in the inset the survival probability of the initial state under ideal evolution, showing that the dynamics in the considered time interval is sizable, namely, the state changes significantly.}\label{fig:Fig9}
\end{figure}

We plot in Fig.~\ref{fig:Fig8} the photon number average $\langle a^\dag a\rangle={\rm Tr}(a^\dag a\rho_{T,I})$ with respect to the states $\rho_T$ (dashed) of the Trotterized evolved dynamics including decoherence sources, and the ideal evolution $\rho_I$ (solid), versus $gt$, with $g=\lambda/\sqrt{N}$. Here $g/\omega=0.5$, which corresponds to the ultrastrong coupling regime. The initial state is the ground state of qubits and bosonic mode of the free Hamiltonian. It can be appreciated how the $N=3$ case (b) has a larger photon emission rate than $N=2$ (a), as corresponds to having a larger number of qubits.

In order to benchmark the generalized Dicke models as well, we performed similar numerical simulations for the pulsed Dicke model, which is the most complex model of the ones proposed in terms of the number of gates. We plot in Fig.~\ref{fig:Fig9} the fidelity $F= {\rm Tr}(\rho_T \rho_I)$ for (a) $N=2$ and (b) $N=3$, of the overlap between the states $\rho_T$ of the Trotterized evolved dynamics including decoherence sources, and the ideal evolution $\rho_I$, versus $g_0t$, with $g_0=\lambda_0/\sqrt{N}$. Here $g_0/\omega=1.5$ and $g_1=2g_0$, with $g_1=(\lambda_0+\lambda_1 \alpha)/\sqrt{N}$. This case corresponds to the deep-strong coupling regime. We depict in the inset the survival probability of the initial state, which is the ground state of qubits and bosonic mode of the free Hamiltonian, showing that the dynamics in the considered time interval is sizable, namely, the initial state significantly evolves. We observe that this state is not preserved as corresponds to this coupling/mode frequency ratio, which includes the counterrotating terms of the Dicke Hamiltonian.

One can appreciate that also in this case the fidelity $F$ decreases from the $N=2$ to the $N=3$ cases, as expected. Moreover, for a reduced number of Trotter steps as compared to the one achievable in the lab~\cite{Langford}, the fidelity remains large with respect to the classical limit during a time evolution with a sizable dynamics, as shown by the inset.

We point out that the pulsed Dicke model is the most complex one proposed, in terms of the number of total gates and number of entangling gates per Trotter step. Therefore, it is expected that the other models introduced, the Fermi-Bose and the biased ones, will behave in general better in the digital-analog quantum simulation, namely, they will require a smaller number of Trotter steps to reach a similar fidelity. 

These numerical simulations are intended to show the feasibility of an experiment with current circuit QED technology. In order to probe the different phases of the system, including the superradiance phase transition, one may modify the protocol to perform a digitized adiabatic evolution~\cite{Barends15a} beginning with an easy to initialize state, and ending with the ground state of the Dicke Hamiltonian or its generalizations. Moreover, as a figure of merit, one may consider the average number of photons in the cavity, $\langle a^\dag a\rangle,$ as a function of time, for different numbers of superconducting qubits, to benchmark the scaling of the collective spontaneous emission rate with $N$, which for a superradiant state is known to grow with $N^2$, see Ref.~\cite{Dicke54}. The quantity $\langle a^\dag a\rangle$ can be efficiently measured with standard techniques, e.g., dual-path methods~\cite{DiCandia}. Having at disposal the flexibility of a DAQS, more exotic phases of matter with partial superradiance, inhomogeneous couplings or free energies, and disorder, can be fully explored with similar techniques. We point out that many of these cases with low or no symmetry will present the sign problem and will be hard for classical computers, such that a digital-analog quantum simulator with a few tens of qubits may solve this kind of dynamics exponentially faster.  Moreover, highly entangled multipartite quantum states may be obtained with these protocols, both in the symmetric~\cite{Bastin} and the general non-symmetric~\cite{Dur,Verstraete,Lamata} subspaces, contributing to complex quantum state generation and studies of quantum correlations in multipartite systems.

\section*{Discussion}
Summarizing, we have proposed a digital-analog quantum simulation of generalized Dicke models with superconducting circuits. We analyze the necessary interactions, and make an estimation of the digital errors, required number of digital steps, and number of gates, comparing them with state-of-the-art experiments on digital quantum simulations with superconducting qubits and circuit QED. Proof-of-principle few-qubit experiments of the Dicke model and its generalizations in a wide range of coupling regions may be performed with current technology. Moreover, improvements in gate fidelity and coherence times may soon allow for the digital-analog quantum simulation of these models involving tens of qubits, which may be significant on the way to quantum supremacy. For fully scaling to several tens or hundreds of qubits, error correction methods may be envisioned for this kind of digital-analog quantum simulator. The digital-analog quantum simulation paradigm is a novel framework that aims at benefitting from the best of current paradigms, digital and analog. Finally, these ideas could be straightforwardly implemented in other quantum platforms, being a prominent example ions in Penning traps~\cite{Bollinger1,Bollinger2}. In this kind of technology, digital-analog quantum simulations of either Dicke or Heisenberg models seem feasible and could in principle involve a large number of qubits, which deserves further analysis.

\section*{Acknowledgements}
The author wishes to acknowledge discussions with I. Arrazola, A. Mezzacapo, J. S. Pedernales, and E. Solano, and support from Ram\'on y Cajal Grant RYC-2012-11391, Spanish MINECO/FEDER FIS2015-69983-P,  UPV/EHU UFI 11/55, EHUA14/04, and Basque Government IT986-16.

\section*{Author contributions}
L. L. envisioned the project, performed all analytical and numerical calculations, analyzed the results, and wrote the manuscript.

\section*{Additional information}

\textbf{Competing financial interests:} The author declares no competing financial interests.

\end{document}